\documentstyle[amssymb,aps,prl,epsfig,twocolumn]{revtex}

\begin{document}
\title{Local vortex nucleation and the surface mode spectrum of large condensates}
\author{J.R. Anglin}
\address{Center for Ultracold Atoms, 26-251\\
Massachusetts Institute of Technology\\
77 Massachusetts Avenue, Cambridge MA 02139}
\date{\today}
\draft
\maketitle

\begin{abstract}
A combination of analytical and numerical approaches obtains the complete
dispersion curve for surface excitations in a condensate held in a plane
linear potential. \ This improvement on previous approximate results yields
an accurate formula for the local Landau critical velocity for vortex
nucleation at the surface of a sufficiently large condensate, which agrees
very well with recent experiments. \ The dispersion curve for surface modes
at a hard wall potential is also presented, for contrast.
\end{abstract}

\pacs{\ 03.75.Fi, 67.40.Db}

\narrowtext

Critical velocities for vortex nucleation in dilute trapped Bose-Einstein
condensates have recently been investigated theoretically \cite
{FCS,FS,SinhaCastin,MF,KPSZ} and experimentally \cite{ENS,MIT,JILA,Oxford}.
\ It is clear topologically and energetically that unstable surface
excitations are the seeds from which vortices grow, and this fixes the
characteristic scales of the problem. \ Quantitative comparison between
theory and experiment, however, requires more than order-of-magnitude
estimates. \ A precise but general theory may be based on the idea that the
Thomas-Fermi (TF) surface of a large condensate enters a `bulk regime' in
which the physics of vortex nucleation is local. \ A merit of this local
theory is that it can be applied to generic condensate flows, which cannot
naturally be described in terms of global surface modes of definite angular
momentum.\ 

When vortices are produced `gently' enough that one can consider them to
grow from a quasi-steady state, the local theory of vortex nucleation
proceeds in three steps. \ The first is determining the condensate density
and velocity field in the vortex-free steady state, as a function of
experimental control parameters such as atom number and stirring frequency.
\ The second is determing the local critical velocity over the TF surface
(which in the presence of stirring beams may be multiply connected), and so
assessing when and where the surface velocity field may be locally
supercritical. \ The third is computing the rate at which supercritical
surface excitations grow into nonlinear vortices. \ The first step is in
general a difficult problem in three-dimensional hydrodynamics, though in
some cases it can be performed analytically. \ The second step is the
subject of this paper. \ The third step again depends on experimental
details, such as kinetics of the thermal cloud, strength of stirring
potential, and timescale over which control parameters are changed. \
Whatever issues arise in steps one and three, the Landau criterion \cite
{Landau} is universal in step two: an energetically unstable surface mode
must appear for vortices to enter the cloud. \ For sufficiently large
condensates, such as some in recent experiments, the local Landau critical
velocity may be determined by numerically solving one universal equation.

Near the Thomas-Fermi (TF) surface of a large trapped condensate, the
trapping potential is approximately linear. \ If the condensate is large
enough, and does not have too extreme an aspect ratio, local physics should
also be insensitive to the curvature of the TF surface. \ So, following Al
Khawaja, Pethick and Smith \cite{AKPS} (hereafter AKPS), one can approximate
a generic TF surface by replacing the actual trapping potential with a
linear ramp $V=Fx$ for constant $F$. \ The equations simplify by rescaling
from physical to dimensionless variables, defining $\vec{x}=\delta ^{-1}\vec{%
x}_{ph}$, $t=\tau ^{-1}t_{ph},\,\psi =\delta \sqrt{8\pi a}\psi _{ph}$, where 
$\delta $ and $\tau $ are the characteristic surface length and time scales 
\begin{equation}
\delta =\left( \frac{\hbar ^{2}}{2MF}\right) ^{1/3}\;,\;\;\tau =\left( \frac{%
2M\hbar }{F^{2}}\right) ^{1/3},
\end{equation}
$M$ is the particle mass, $a$ is the s-wave scattering length, and $\psi
_{ph}$ is the macroscopic wavefunction. \ (The chemical potential factor $%
e^{-i\mu t_{ph}}$ has been extracted from $\psi _{ph}$, which is normalized
so that $\left| \psi _{ph}\right| ^{2}$ is the density of particles in the
condensate.) \ In the dimensionless variables the Gross-Pitaevskii equation
then reads 
\begin{equation}
i\partial _{t}\psi =-\nabla ^{2}\psi +x\psi +\left| \psi \right| ^{2}\psi \;,
\end{equation}
where the origin of $x$ is at the TF surface. \ Setting $\partial _{t}\psi
=0 $ determines the stationary background solution $\psi =f(x)$, where $f$
is taken to be real. \ It may be found to good accuracy numerically \cite
{DPSLPS}.

The dispersion relation for surface excitations may be found by solving the
Bogoliubov equations for small perturbations $\psi \rightarrow f+\eta
\,\delta \psi ,$ with infinitesimal $\eta $\ and 
\begin{eqnarray}
\delta \psi &=&u(x)\;e^{i(\vec{k}\cdot \vec{x}_{ph}-\omega t_{ph})}+v^{\ast
}(x)\;e^{-i(\vec{k}\cdot \vec{x}_{ph}-\omega t_{ph})}  \nonumber \\
&=&u(x)\;e^{i(qy-\Omega t)}+v^{\ast }(x)\;e^{-i(qy-\Omega t)}.
\end{eqnarray}
Here the excitation's wave vector is taken to lie along the $y$ axis, and
the dimensionless frequency $\Omega =\omega \tau $\ and wave number $%
q=k\delta $ are defined. \ Note that far outside the TF surface, for $%
x\rightarrow \infty ,$ $f(x)\propto e^{-\frac{2}{3}x^{3/2}}\rightarrow 0$,
and hence 
\begin{eqnarray*}
&\because &\lim_{x\rightarrow \infty }\delta \psi \propto e^{i(qy-\Omega
t)}\exp [-\frac{2}{3}x^{3/2}+(\Omega -q^{2})x^{1/2}] \\
&\therefore &\lim_{x\rightarrow \infty }\psi \propto e^{-\frac{2}{3}x^{3/2}}%
\left[ 1+\eta e^{(\Omega -q^{2})x^{1/2}}e^{i(qy-\Omega t)}\right] .
\end{eqnarray*}
Since it turns out that $\Omega >q^{2},$ this means that the growth of $\eta 
$ does not just lead to vortex formation, but actually {\em is} the approach
of vortices to the TF surface from infinity. \ 

It is convenient to solve for the dispersion relation $\Omega (q)$ by
iterating the Bogoliubov equations to obtain the fourth order eigenvalue
problem for $s(x)=u-v$: 
\begin{eqnarray}
\Omega ^{2}s &=&(\hat{H}_{3}+q^{2})(\hat{H}_{1}+q^{2})s  \label{eigen} \\
\text{where \ }\hat{H}_{n} &\equiv &-\partial _{x}^{2}+x+nf^{2}(x)\;.
\end{eqnarray}
Discretizing $x$ and representing the second derivatives with finite
differences converts this fourth order ODE directly into a matrix
diagonalization problem, which can be solved numerically. \ Taking $\Omega
(q)$ to be the square root of the lowest eigenvalue $\Omega ^{2}$ for given $%
q$, one finds the function shown in Figure 1. \ (For any $q$ there is
obviously a discrete spectrum of $\Omega ^{2}$, but the critical velocity is
set by the lowest branch.) \ Numerical solution becomes difficult at small $%
q $, where even the lowest lying eigenstate $s(x)$ extends far enough into
the condensate that a large grid is needed to contain it; but fortunately
the problem remains easily tractable until the numerical curve has already
converged onto the $\sqrt{2q}$ asymptotic behaviour as $q\rightarrow 0$,
found analytically by AKPS \cite{AKPS}. \ At large $q$, simple perturbation
theory implies that $\Omega (q)=q^{2}+E_{g2}+{\cal O}(q^{-2})$, where $%
E_{g2}\doteq 1.17$ is the lowest eigenvalue of $\hat{H}_{2}$. \ (This is the
same as the `Popov approximation' of neglecting the anomalous average in the
Bogoliubov equations for $u$ and $v$.) \ Figure 1 shows that the exact
dispersion curve approaches these asymptotic forms closely for $q\lesssim
0.3 $ and for $q\gtrsim 1.5$. \ 

For comparison, Figure 1 also shows the approximation of AKPS \cite{AKPS},
which in our dimensionless units is 
\begin{equation}
\Omega _{AKPS}^{2}=2q+4q^{4}\left[ 0.15-\ln \left( q\right) \right] \;,
\label{KPS}
\end{equation}
and was derived to give the first corrections at small $q$ beyond the
leading term $\Omega ^{2}\sim 2q$. \ According to Figure 1, it clearly does
this very well, tracking the exact dispersion curve closely for $q\lesssim
0.6$. \ This approximation was never intended to be accurate at larger $q$,
however, and it obviously becomes pathological for $q\gtrsim 1$; its
application to vortex nucleation in \cite{FCS}\ is therefore only
qualitatively accurate. \ The recent \cite{MF}, on the other hand, is based
on the high $q$ regime, in the sense that it fits numerical data to the
functional form $q^{2}+$ const., and so underestimates $v_{c}$ by about
15\%, because this form is not really valid near $q_{c}$. \ For comparison
with (\ref{KPS}), in the range $0\leq q\leq 2$ the numerical result shown in
Figure 1 is indistinguishable, on the scale of the plot, from 
\begin{equation}
\Omega _{0-2}^{2}=2q+1.35\,q^{3}+0.711\,q^{4}.
\end{equation}
This formula becomes inaccurate at $q\gg 2$, where it is better to use the
asymptotic Popov result.

\epsfig{file=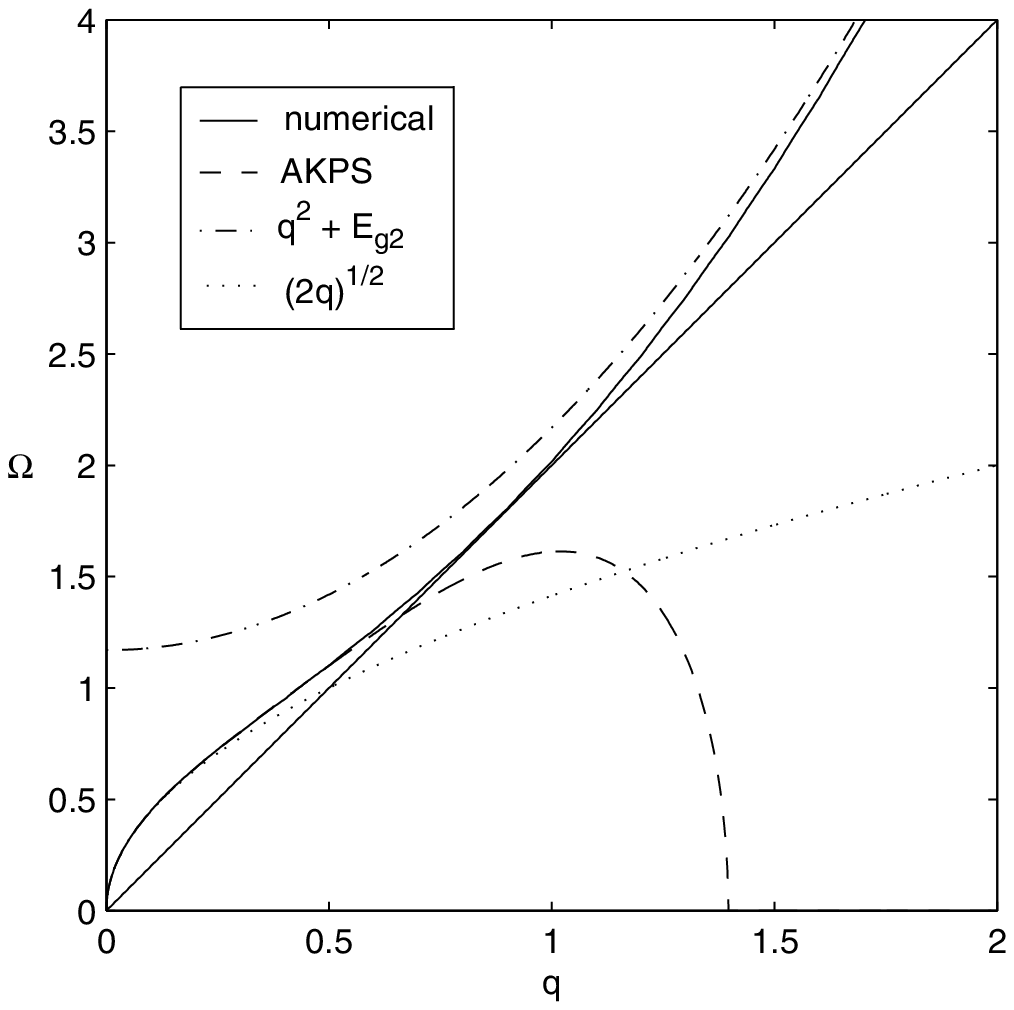,width=.45\textwidth} \vspace*{3mm} 
\begin{minipage}{.475\textwidth}\small
\indent
{\bfseries Figure 1. }  Dimensionless surface mode dispersion relation 
$\Omega(q)$ (solid curve).  This curve is numerical for $q>0.1$; for $q<0.1$ it is simply 
$\sqrt{2q}$.  The tangent to $\Omega (q)$ through the origin has slope 2.  The 
approximation of Al Khawaja, Pethick and Smith is shown for comparison. 
Dotted curves show the asymptotic behaviours $\sqrt{2q}$ and 
$q^{2}+E_{g2}$ (see text).  
\end{minipage}
\ 

The main result which we can derive from our numerical knowledge of $\Omega
(q)$ is an accurate value for the Landau critical velocity due to surface
excitations, which in units of $\delta /\tau $ is simply the slope of the
tangent to $\Omega (q)$ through the origin. \ Numerically this slope is
indistinguishable from 2. This leads to the expression in physical units 
\begin{equation}
v_{c}=\left( \frac{2\hbar F}{M^{2}}\right) ^{1/3}=\frac{\hbar }{M}\,\delta
^{-1}.  \label{critvel}
\end{equation}
\ This formula can immediately be obtained as an order of magnitude
estimate, of course; but the fact that it is precisely accurate is
nontrivial. \ Despite its simple appearance, I have been unable to derive
this result analytically, and it may not actually be exact; but it is
accurate at least to within a few tenths of one percent. \ Another result
which we obtain is the wavenumber $q_{c}$ at which $\Omega (q)/q$ is
minimized, which is approximately $q_{c}\doteq 0.89$. \ This number is close
to $2^{-1/6}$; but since $\Omega (q)$ has rather small curvature near $%
q=q_{c}$, the numerical uncertainty in $q_{c}$ is too large to make this
identification more than a conjecture. \ 

The applicability of the local surface critical velocity (\ref{critvel}) to
recent experiments may be assessed as follows.\ Since $\delta $ is the
characteristic depth to which surface modes of $q\sim 1$ penetrate the
condensate, the bulk regime requires $\delta /R=\varepsilon \ll 1$ (where $R$
is the TF radius of the condensate at its rotational equator) so that
surface modes are indeed confined to within the range over which the
harmonic potential is accurately linear. \ Since $\pi \delta $ is also the
characteristic surface wavelength of the critical excitations, neglecting
the curvature of the TF surface requires that $\delta $ is much smaller than
either radius of curvature. \ In prolate traps rotated about the long axis,
the shortest radius of curvature is just $R$; but\ in oblate traps the
curvature in the axial ($z$) direction is greater, and we need the more
stringent condition $\varepsilon =(\omega _{z}^{2}\delta )/(\omega
_{r}^{2}R)\ll 1$.\ \ 

For a rotationally symmetric harmonic trap, the quantity $(q_{c}R/\delta -1)$%
\ may be compared to the multipolarity $l_{c}$ of the rotationally symmetric
surface mode, and the surface critical velocity can be converted into the
dimensionless critical rotation frequency as a fraction of radial trapping
frequency, given in the Thomas-Fermi limit of large particle number $N$\ by 
\begin{equation}
\chi =\frac{\omega _{c}}{\omega _{\rho }}=2^{1/3}\left( \frac{\hbar \omega
_{\rho }}{M\omega _{z}^{2}a^{2}}\right) ^{\frac{1}{15}}\left( 15\,N\right)
^{-\frac{2}{15}}.  \label{anything}
\end{equation}
\ For experiments reported to date, computing $F$ using the TF density
profile for the reported numbers of atoms yields

\begin{center}
\begin{tabular}{|c|c|c|c|c|c|}
\hline
Experiment & $N$ & $\varepsilon $ & $q_{c}R/\delta $ & $\chi $ & $\chi
_{obs} $ \\ \hline
MIT stiff \cite{MIT} & $1.5\times 10^{7}$ & $0.045$ & $19.6$ & $0.30$ & $%
0.29 $ \\ \hline
MIT weak\cite{MIT} & $5\times 10^{7}$ & $0.030$ & $29.8$ & $0.24$ & $0.27$
\\ \hline
ENS \cite{ENS} & $3\times 10^{5}$ & $0.11$ & $8.4$ & $0.46$ & $0.65$ \\ 
\hline
JILA \cite{JILA} & $6\times 10^{6}$ & $0.12$ & $30.7$ & $0.24$ & $0.3(5\pm
3) $ \\ \hline
Oxford \cite{Oxford} & $2\times 10^{4}$ & $0.72$ & $9.9$ & $0.42$ & $0.56+$
\\ \hline
\end{tabular}
\end{center}

\noindent The values of $\varepsilon $ (which include oblateness factors of
4 and 8 for the JILA and Oxford traps respectively) indicate that the local
surface theory should describe the MIT experiments very well, and should
apply to the JILA and ENS experiments with fair accuracy, but is not
expected to fit the measurements at Oxford (which were made at various
eccentricities, not reflected in the table, and to which the global
hydrodynamic analysis of \cite{SinhaCastin, KPSZ} is relevant instead). \ 

The local theory actually agrees even better with the MIT experiments than
the table indicates, because with the stirring beam applied the MIT traps
were significantly asymmetric, and for the weak configuration even
anharmonic. \ Hydrodynamics in the weaker trap is therefore not analytically
tractable, but the stiff MIT\ trap remains harmonic with the stirring beam
applied, and so the hydrodynamic solution for a rotating harmonically
trapped condensate \cite{quadrupole} yields the non-perturbative density and
velocity fields, in the co-rotating frame. \ One can use this solution
straightforwardly, for a given trap eccentricity and rotation frequency, to
compute the local condensate velocity $v(\theta ,\phi )$ over the whole TF
surface, as well as the local surface force $F(\theta ,\phi )$ (taking
centrifugal effects into account), and hence the local critical velocity $%
v_{c}(\theta ,\phi )$. \ The results show that the local flow velocity first
reaches the local critical velocity at the poles of the cloud's shortest
axis, so that vortices will first form in these surface regions. \ One also
finds that local criticality is reached in the stiff MIT trap at a slightly
lower $\chi $ of $0.285$. \ (Although angular momentum is not conserved in
the asymmetric trap, the angle subtended by $2\pi \delta /q_{c}$ at the
vortex nucleation radius (smallest semi-axis) yields the effective $l\doteq
18$ cited in \cite{MIT}.)

The excellent agreement with the MIT experiments is likely due not only to
the lower values of $\varepsilon $, but also to the facts that the MIT
experiments were most sensitive in detecting vortices (due to condensate
size and a `slicing' technique to improve visibility), and applied strong
stirring fields (so that the timescales for vortex nucleation were probably
shorter for a given degree of supercriticality). \ As suggested in \cite{ENS}%
, the nucleation time in the weakly anisotropic ENS trap may be too long for
visible vortices to be created, except in the large (and unstable \cite
{SinhaCastin}) velocity fields generated by rotations near the quadrupole
resonance. \ Similarly, for the JILA system the $\chi =0.24$ given by (\ref
{anything}) is significantly lower than the lowest value $0.32$ at which
vortices may have been detected \cite{sumrule}.\ \ This discrepancy is
enough to motivate further development of the local critical velocity
theory, to determine whether ${\cal O}(\varepsilon )$\ corrections due to
surface curvature may have large constant factors. \ It is also worth
noting, however, the significant experimental uncertainty due to the
indirect detection of small numbers of vortices through changes in the
condensate aspect ratio \cite{JILA}. \ And since vortex nucleation in the
JILA experiments is through equilibration with the rotating thermal cloud,
one can estimate the rate of surface instability growth using quantum
kinetic theory, and show that vortex nucleation is slow until significantly
supercritical velocities are reached. \ (When the thermal cloud's mean
velocity past the TF surface just exceeds $v_{c}$ \cite{QKT}, the highest
gain over loss is for the mode with $q=q_{c}$, and is proportional to $%
(1-\exp [-2\beta q_{c}(v-v_{c})])$, where $\beta $ is the thermal cloud's
inverse temperature in our dimensionless surface units. \ Since experimental
temperatures of $T\sim 100$nK are much higher than the temperature
corresponding to the surface critical velocity, this again indicates that
the timescale for vortices to form may be quite long until the critical
rotation frequency has been significantly exceeded.) \ Further study of such
timescales is needed, but it appears that the JILA data do not demand any
vortex nucleation mechanism other than surface mode excitation by the `wind'
of the rotating thermal cloud.

More recent observations at MIT \cite{MIT2} using a narrow, stick-like
stirring beam have obtained $\chi $ as low as $0.08$, where Eqn. (\ref
{anything}) predicts $0.24$. \ As in the rotating stiff trap, however, the
local critical velocity must be compared to the actual local fluid velocity
created over the entire TF surface, including the surface at the stirring
beam. \ Detailed analysis of the 3D flow induced by the Gaussian beam will
require substantial numerical effort, and is beyond the scope of this paper;
but simplified simulations show that fluid velocities near Gaussian beams
can easily be many times the velocities of the beams themselves. \ (The
well-known hydrodynamic speed-up factor of 2 for a hard cylindrical stirrer
is enhanced because the fluid density drops as it penetrates the beam.) \ Of
course if the beams are too narrow in relation to their strength, the ratio
of surface depth to curvature radius on the TF surface may not be
sufficiently small for the zeroth order theory presented here to be
accurate. \ On the other hand, if the $\delta $ on the beam surface is
actually smaller than the healing length $\xi =(4\pi a|\psi
_{ph}|^{2})^{-1/2}$ near the beam, the linearized theory will also fail, as
the beam potential approaches the limit of a hard wall. \ The surface mode
spectrum of a hard wall can also be computed, if we can again neglect
curvature. \ In this case we have an analytic solution for the stationary
background $\psi \propto \tanh x/\xi $, and a multiple scale analysis valid
at long wavelength yields lowest Bogoliubov frequencies $\omega _{\xi
}\equiv (2M\xi ^{2}/\hbar )\omega =2k\xi +{\cal O}[(k\xi )^{5}]$. As shown
in Fig. 2, the result is that the surface mode dispersion curve $\omega
_{\xi }(k\xi )$\ for a hard wall lies below the bulk dispersion curve, but
not by enough to lower the critical velocity below the local speed of bulk
sound.\ 

\epsfig{file=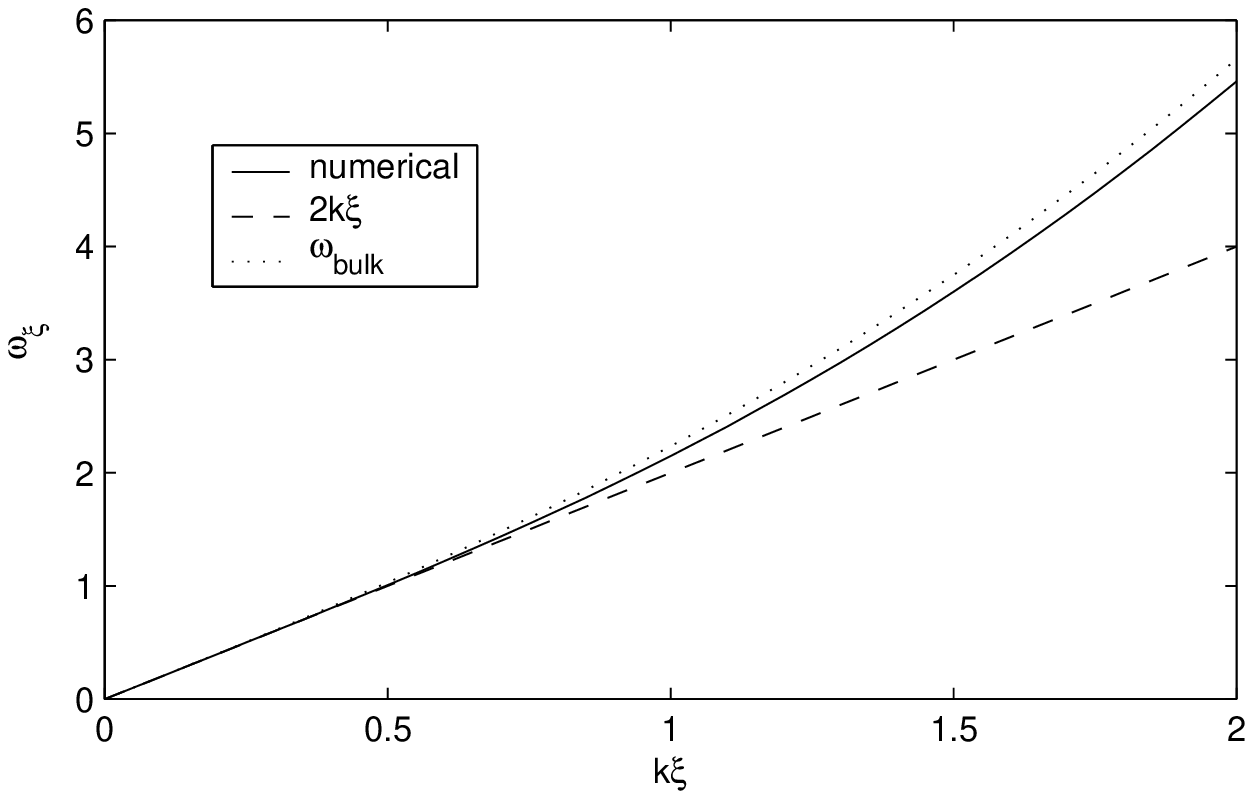,width=.45\textwidth} \vspace*{3mm} 
\begin{minipage}{.475\textwidth}\small
\indent
{\bfseries Figure 2.} Hard wall surface mode dispersion relation 
$\omega_\xi(k\xi)$ (solid curve). \ Results are numerical only for $q>0.1$;
below this point the solid curve is simply $2k\xi$.  For all $k$, the surface mode dispersion 
curve lies between $2k\xi$ and the bulk dispersion curve $k\xi\sqrt{4+k^2\xi^2}$.  
\end{minipage}

To summarize, in condensates of at least moderate size vortex nucleation at
TF surfaces is governed by the local Landau critical velocity obtained from
the dispersion relation for Bogoliubov surface modes. \ This velocity is $%
\hbar /(M\delta )$ to high accuracy (not just of that order). \ Experiments
at MIT support this theory well. \ Failure to see vortices at such low
velocities elsewhere is likely due to long nucleation times. \ For very
abrupt potentials, the linear theory breaks down and the local surface mode
critical velocity approaches the local speed of bulk sound.

It is a pleasure to thank J.R. Abo-Shaeer, C. Raman, and W. Ketterle for
valuable discussions. \ This work was supported by the NSF through its
grants to the Institute for Theoretical Atomic and Molecular Physics and to
the Center for Ultracold Atoms.


\begin{references}
\bibitem{QKT}  Imposing conservation of energy and momentum in collisions
that generate surface excitations, one obtains that a surface mode of
dimensionless wavenumber $q$ can only be excited by incident thermal atoms
whose momentum component in the direction of the surface mode is $%
q_{I}=[\Omega (q)+q^{2}]/(2q)$. \ The same surface mode can only be
de-excited by incident thermal atoms of $q_{I}=[\Omega (q)-q^{2}]/(2q)$. \
For a Boltzmann distribution of thermal atoms $P\propto \exp [-\beta
(q_{I}-v/2)^{2}]$, therefore, the ratio between rates of gain and loss can
only exceed unity for any surface mode $q$ if $v>v_{c}$. \ So even though
the threshold velocity for an incident atom to excite surface modes at $%
q_{c} $\ is higher than $v_{c},$ due to atomic recoil, the critical velocity
for a thermal cloud of non-interacting atoms is still exactly $v_{c}$.
\end{references}
\end{document}